\PassOptionsToPackage{dvipsnames}{xcolor}
\documentclass[10pt,conference]{IEEEtran}
\IEEEoverridecommandlockouts

\usepackage[all]{nowidow}
\usepackage{float}
\usepackage{xcolor}
\usepackage{subcaption}
\usepackage{hyperref}
\usepackage{acronym}
\usepackage{xparse}
\usepackage{booktabs}
\def\BibTeX{{\rm B\kern-.05em{\sc i\kern-.025em b}\kern-.08em
    T\kern-.1667em\lower.7ex\hbox{E}\kern-.125emX}}

\NewDocumentCommand{\codeword}{v}{%
\texttt{\textcolor{orange}{#1}}%
}

\usepackage[capitalise]{cleveref}
\crefformat{section}{\S#2#1#3} 
\crefformat{subsection}{\S#2#1#3}
\crefformat{subsubsection}{\S#2#1#3}

\usepackage{todonotes}
\begin{document}
\title{A Crowdsensing Approach for Deriving Surface Quality of Cycling Infrastructure}

\author{\IEEEauthorblockN{Ahmet-Serdar Karakaya}
\IEEEauthorblockA{\textit{TU Berlin \& ECDF} \\
Berlin, Germany \\
ask@mcc.tu-berlin.de}
\and
\IEEEauthorblockN{Leonard Thomas}
\IEEEauthorblockA{\textit{TU Berlin} \\
Berlin, Germany \\
leth@mcc.tu-berlin.de}
\and
\IEEEauthorblockN{Denis Koljada}
\IEEEauthorblockA{\textit{TU Berlin} \\
Berlin, Germany \\
deko@mcc.tu-berlin.de}
\and
\IEEEauthorblockN{David Bermbach}
\IEEEauthorblockA{\textit{TU Berlin \& ECDF} \\
Berlin, Germany \\
db@mcc.tu-berlin.de}
}

\maketitle

\begin{abstract}
Cities worldwide are trying to increase the modal share of bicycle traffic to address traffic and carbon emission problems.
Aside from safety, a key factor for this is the cycling comfort, including the surface quality of cycle paths.
In this paper, we propose a novel edge-based crowdsensing method for analyzing the surface quality of bicycle paths using smartphone sensor data:
Cyclists record their rides which after preprocessed on their phones before being uploaded to a private cloud backend.
There, additional analysis modules aggregate data from all available rides to derive surface quality information which can then used for surface quality-aware routing and planning of infrastructure maintenance.\end{abstract}


\section{Introduction}
\label{sec:intro}
Cities all over the world aim to increase the modal share of bicycle traffic, e.g., to address emission problems, frequent traffic jams, but also to improve the citizens' health through more daily activity.
Key factors for this are traffic safety, topography but also the comfort level of cyclists~\cite{Nyberg1996Road,gadsby2022understanding,karakaya2020simra}.
All these factors are affected by the availability and quality of cycling infrastructure which needs to be monitored and maintained continuously.
Even monitoring of surface quality alone, however, can be challenging due to the sheer dimensions of existing infrastructure.
Germany's capital Berlin, for example, has around 2,300 km of bicycle tracks as of March 2023, i.e., manually inspecting cycling infrastructure to monitor surface quality is infeasible.

To address this problem, we propose to use a crowdsensing approach in which cyclists record their daily rides using a smartphone app.
Using the built-in motion sensors, the surface quality (or rather the lack of) can be measured as vibrations or bumps.
In a second step, we can then combine data from multiple rides to derive an estimate for the surface quality.
This way, monitoring of surface quality can be automated to a high degree at little cost and the resulting data can be used for maintenance planning or surface quality-aware routing.
Especially for highly frequented cycling tracks, surface quality problems can be detected quickly.

While there are other projects studying the surface quality of cycling infrastructure through crowdsourcing (e.g., Luedemann et al.~\cite{luedemann2022bikevibes}), our work is unique and novel because alternative approaches (i) either focus on categorizing the surface \emph{type} (e.g., cobblestones) where we focus on the surface \emph{quality} (i.e., the level of roughness) or (ii) have strict assumptions on phone positioning and other properties where we rely on a ``wisdom of the crowds'' strategy to filter out noise.
The latter is also an advantage because, unlike related approaches, which use a dedicated app for measuring the road surface quality, thus, potentially forcing cyclists to run multiple apps, our approach can easily be retrofitted to existing apps and can even be used to analyze already stored datasets.

For this, we make the following contributions:
\begin{itemize}
	\item We present a data processing pipeline for deriving surface quality which combines signal processing with geographical clustering techniques in the form of edge-based preprocessing on the phone and cloud analytics (\cref{sec:pipeline}).
	\item We describe how we integrate this data processing pipeline in the existing crowdsensing project SimRa~\cite{karakaya2020simra} which up to now has only focused on traffic safety(\cref{sec:pipeline}).
	\item We describe how the resulting data can be used to increase the comfort of cyclists through surface quality-aware routing and how the data can be exposed to city administrations(\cref{sec:use}).
	\item We evaluate our approach by analyzing the SimRa dataset~\cite{dataset_simra_set1,dataset_simra_set2,dataset_simra_set3} and comparing it on-site conditions in eight streets with different surface quality (\cref{sec:eval}).
	\item We discuss to which degree our approach can automate surface quality monitoring and how additional sensors could possibly help to improve data quality (\cref{sec:disc}).
\end{itemize}

\section{Scenarios and Constraints}
\label{sec:scenarios}
The idea to use smartphones and crowdsensing for tracking surface quality is not new.
For this reason, we use this dedicated section to briefly discuss the specific goals of our work and the unique constraints of the scenarios we target.
We do this to clarify why existing approaches, which we discuss in the next section, cannot solve the problem we target.

\begin{figure*}[t]
    \centering
    \subfloat[Smooth cobblestones]{\includegraphics[width=\columnwidth]{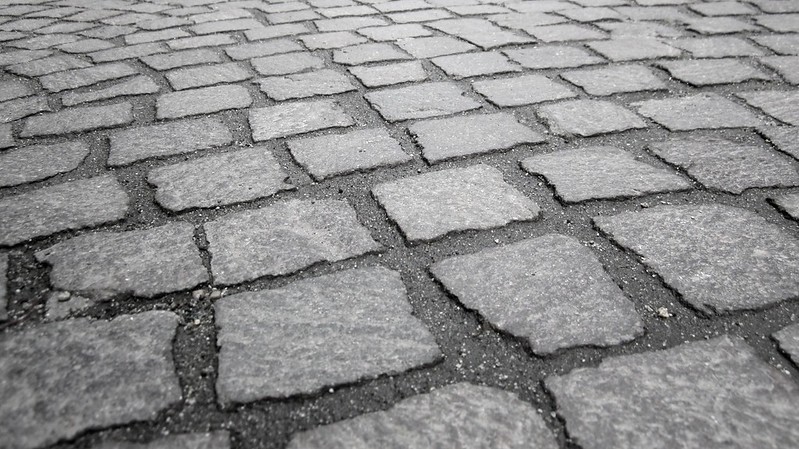}}
    \hfill
    \subfloat[Rough asphalt]{\includegraphics[width=\columnwidth]{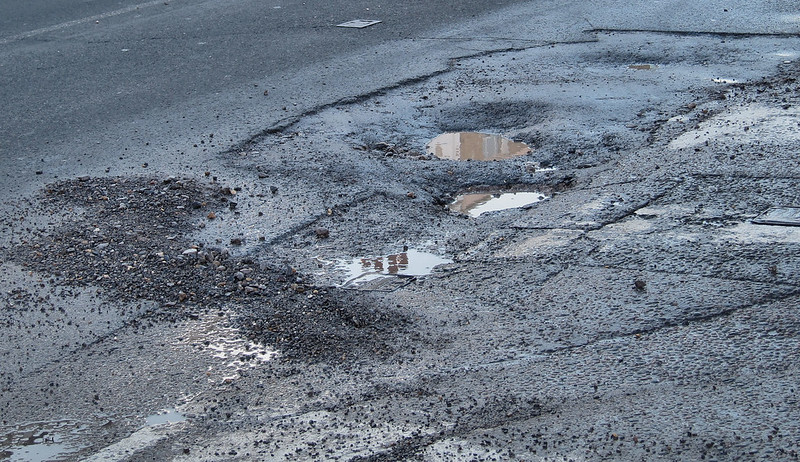}}
    \caption{%
        Smooth cobblestones can provide a good surface quality while bad asphalt can result in a bumpy ride (Source: Flickr.com).
    }%
    \label{fig:cobblestone_vs_asphalt}
    \vspace{-.5em}
\end{figure*}

First, our work focuses on the experience in terms of ``bumpiness'' for cyclists.
Therefore, we try to quantify the roughness of a road surface and not its type.
As an example, a ride on asphalt will usually be smoother than one on a cobblestone road.
In practice, though, an asphalt road might have lots of (small) potholes and repair patches while a cobblestone road might use relatively flat (instead of rounded) stones  with mostly filled-in gaps between the stones.
See \cref{fig:cobblestone_vs_asphalt} for an example.
In such a setup, a ride on the cobblestone road can be much smoother.
Quantifying the surface \emph{quality} and not the surface \emph{type} will thus yield different results -- a large body of related work is hence not applicable to the problem we target.

Second, due to the width of bicycle tires, a single ride on a street segment will only cover a very small percentage of the surface.
Furthermore, cyclists are likely to swerve around the worst potholes and tree roots.
Both aspects combined show that it is crucial to base the roughness measurements on a large number of rides, thus, following a ``wisdom of the crowds'' approach.
We hence have to attract a large user group -- this has a number of implications.
\begin{enumerate}
	\item The approach needs to go easy on the phone's resources (battery, data transmission, compute power, app size) as users will otherwise uninstall or not use the corresponding app. As a result, the approach will have to prefilter data on the phone but is unlikely to run on the phone completely, machine learning-based approaches may be problematic, and the approach cannot expect high resolution raw data.
	\item The approach should not have physical setup requirements. Some cyclists mount their phones on the handlebar (which usually will yield the best results for using accelerometer sensors), others keep it in pockets, backpacks, etc. Any approach that \emph{requires} cyclists to use a certain setup will deter a large number of potential users. As an implication, many approaches that work under lab conditions will not work on the street.
	\item The approach needs to easily integrate into existing cycling apps. As we have seen in SimRa~\cite{karakaya2020simra}, many cyclists will not use more than one cycling app in parallel. To maximize the potential user base, the approach hence needs to be designed in a way that it can easily be integrated into existing cycling apps. In general, this means that the approach needs to run as an independent background job and may not involve manual labeling and similar activities. They may also be subject to privacy-related restrictions.
\end{enumerate}

The approach we present in this paper was designed around these constraints.
We retrofitted it to the SimRa app~\cite{karakaya2020simra} but it could easily be integrated as a plugin in, e.g., Strava or BikeCitizens.
As a soft constraint, we tried (and succeeded) to design our approach in a way that it can also process our existing datasets~\cite{dataset_simra_set1,dataset_simra_set2,dataset_simra_set3}.

\section{Background and Related Work}
\label{sec:rw}
As described, we propose a smartphone-based crowdsensing approach that uses accelerometer sensors for quantifying surface quality.
In this section, we give an overview of alternative approaches from related work focusing on surface quality measurement (\cref{subsec:surface-rw}) as well as similar approaches from other domains (\cref{subsec:edge-rw}).

\subsection{Measuring Surface Quality}%
\label{subsec:surface-rw}
Surface quality is usually quantified based on the International Roughness Index (IRI), e.g.,~\cite{sayers1986The}.
Traditionally, this was done using so-called profilographs (approximately resembling a large ladder with wheels) which are towed by a car or pushed manually.
A study from the 1980s~\cite{cumbaa1986road} found that they often break down and are difficult to maneuver in narrow streets which renders them infeasible for measuring the quality of bike lanes.
Furthermore, such measurement are very personnel-intensive which makes it unrealistic to apply them on a broad scale.

Taniguchi et al.~\cite{taniguchi2015evaluation} detect road hazards such as debris, potholes, or bumps.
For this, they attach an ultrasonic distance sensor to a bike handlebar, scanning the ground in front of the bicycle.
While the approach can warn cyclists about incoming bumps on the road ahead in real-time, this approach does not scale for city-wide analysis due to the sheer number of sensors needed.

Peng et al.~\cite{peng2019road} follow the same goal, using motion sensors instead of distance measurements.
Data is analyzed offline using a classifier which can identify asphalt, pebbles, and very bumpy underground.
In contrast to our work, their emphasis is on identifying specific surface \textit{types} rather than the surface \textit{quality}.
Also, their approach again requires dedicated hardware.

Zhou et al.~\cite{zhou2022smartphone} try to detect manhole covers.
For this, they analyze a video stream from a bike handlebar-mounted smartphone camera using a convolutional neural network.
While their approach is similar to ours regarding hardware, constantly recording video means high power consumption.
Furthermore, the phone needs to be mounted in an awkward angle which makes it impractical for every day use.

Luedemann et al.~\cite{luedemann2022bikevibes} use a smartphone app to record accelerometer data as an indicator for surface quality.
Their approach, however, relies on single rides and has no concept of aggregating data from multiple rides.

Similar to us, Yamaguchi et al.~\cite{yamaguchi2015simple} want to measure the surface quality.
To achieve highly precise results, they combine the smartphone motion sensors with a cyclometer.
This approach will always give more precise results than our approach but again requires dedicated hardware which renders a city-wide usage infeasible.

Beyond these, there are several car-centric approaches which either use dedicated hardware, e.g.,~\cite{paper_chen_crsm}, have very specific phone placement requirements, e.g.,~\cite{paper_daraghmi_surface_crowdsourcing}, or focus on detecting the transients, i.e., individual pot holes, e.g.,~\cite{paper_alam_surface_transients,paper_li_surface_transients,paper_lima_surface_transients}.

\subsection{Edge-assisted Data Analysis}%
\label{subsec:edge-rw}
In previous work~\cite{karakaya2020simra,karakaya2022cyclesense,temmen2022crowdsourcing}, we have described the SimRa project on tracking near miss incidents in bicycle traffic using the built-in motion sensors of off-the-shelf smartphones.
During rides, SimRa records a GPS trace at 1/3Hz and the motion sensors at 50Hz.
To preserve bandwidth and save storage space, the motion sensor data is on the phone downsampled by calculating a moving average with a window size of 30 and then keeping only every fifth value.
After the ride, users can crop and annotate the recorded route before upload.
In this paper, we build upon the project and integrated our data processing pipeline into the existing SimRa app, using the same data measurements.

Other edge-assisted data analysis work includes Mei et al.~\cite{mei2017ultraviolet}, who measure UV radiation based on smartphone cameras and crowdsensing,
Cao et al.~\cite{cao2015fast} who use motion sensors to detect strokes in patients falling to the ground, and
Pham et al.~\cite{pham2015a} implement a smart parking system by equipping parking spots with RFID chips to track their occupancy state.
There are also multiple publications on placing different components of an edge-to-cloud data processing pipeline in various use cases, e.g.,~\cite{paper_lujic_intrasafed5g,paper_pfandzelter_zero2fog,hattab2019optimized}.

\section{Data Processing Pipeline}
\label{sec:pipeline}
In this section, we start by giving a high-level overview of our data processing pipeline for deriving surface quality \cref{sec:overview} before describing the individual steps in the process (\cref{sec:data_acq} to \cref{sec:surface_quality}).

\subsection{Overview}
\label{sec:overview}
\begin{figure}
    \centering
    \includegraphics[width=\columnwidth]{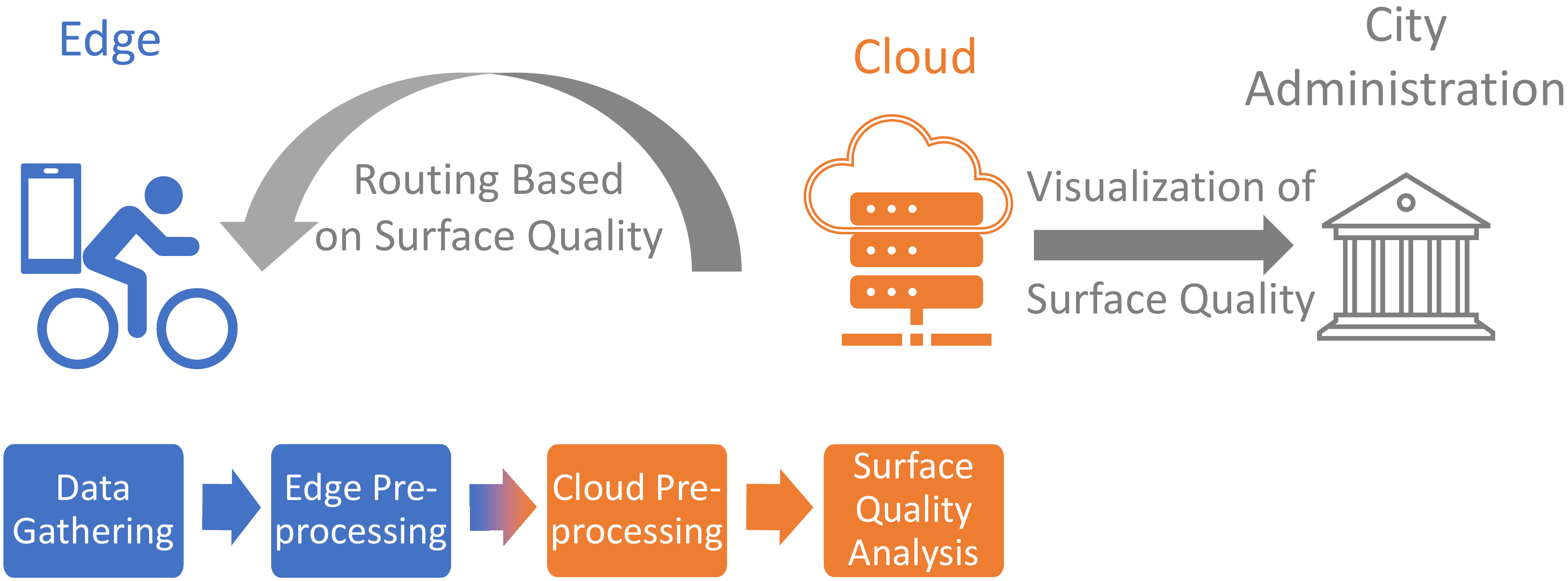}
    \caption{%
        Overview of our surface quality analysis pipeline.
    }%
    \label{fig:overview}
\end{figure}
The tasks of our data processing pipelines are distributed over the edge (i.e., on the smartphone) and the cloud, see also \Cref{fig:overview}.
After collecting data in the SimRa app using the built-in motion sensors, data is preprocessed locally before upload to our backend servers.
There, additional preprocessing steps are executed before the actual surface quality analysis.

The parts of the preprocessing which reduce the amount of data need to be run on the edge to preserve user privacy and to reduce bandwidth consumption.
The data cleaning parts are too compute-intensive and are executed in the cloud to reduce power consumption on the phone.

\subsection{Data Gathering}%
\label{sec:data_acq}
The data is based on crowdsourcing and is gathered by citizen scientists voluntarily using the SimRa app~\cite{karakaya2020simra,temmen2022crowdsourcing}.
While SimRa was originally created for detecting near miss hotspots in bicycle traffic~\cite{karakaya2022cyclesense}, the ride and near-miss incident data it provides can also be used for other goals, such as creating a bicycle model in traffic simulation~\cite{karakaya2022realistic,paper_karakaya_simra_sumo} or, as we do here, for analyzing the surface quality of the bike lanes.
As we planned the secondary use of surface quality analysis early on, SimRa also collects detailed accelerometer data as described in \Cref{subsec:edge-rw} 
While using crowdsensing in the SimRa way can easily scale out to large numbers of users, thus resulting in large volumes of data, it tends to produce noisy data which requires preprocessing before the actual data analysis.

\subsection{Preprocessing on the edge}%
\label{sec:preprocessing_edge}
To reduce the volume of data recorded and preserving privacy of users, we run two preprocessing tasks on the phone.
First, during the ride, we downsample the accelerometer data.
Originally queried at 50Hz, we calculate a moving average of length 30 and then take every fifth value.
With this, we aim to catch as many peaks as possible while keeping data volumes at a reasonable level.
Second, users can crop their rides before the upload to hide their origin or destination.

\subsection{Preprocessing in the Cloud}%
\label{sec:preprocessing_cloud}
The preprocessing in the cloud has two main goals: cleaning noisy data and calculating a value that represents the smoothness of the ride.
For the first goal, we cut off ten seconds each at the start and end of the ride to remove noise resulting from users (un)mounting their bikes or putting the phone in a pocket.
Ideally, this would of course already be done on the phone but could not be done here since the main use case of SimRa requires this data, i.e., we would have had to upload the data twice.
We also remove stops, i.e., only the parts of the ride which have a minimum speed of 5 km/h and a minimum duration of 1 minute are considered.
Such low speeds usually occur, when the bicycle is being pushed by the cyclist rather than being ridden or when stopping at traffic lights.
Even if the cyclist should manage to ride so slowly, the ride will be prone to unsteady motions, leading to noisy data.
This also removes parts where users forgot to stop the recording after their ride.

For the second goal, we normalize the accelerometer data by calculating the mean of the three moving variances (window size 10) of the axes X, Y and Z.
This preserves information about amplitude but disregards the direction which is highly affected by the position of the phone.
Thus, we obtain a time series of single values representing the bumpiness of the surface in each in-motion ride segment.

\subsection{Surface Quality Analysis}%
\label{sec:surface_quality}
Each cyclist creates a different data track on a given road, and without calibration, it is not possible to accurately analyze each track individually.
Ideally, we would at least use cyclist-specific profiles (i.e., per cyclist aggregates) which, however, are not available due to privacy reasons (rides are pseudonymized individually~\cite{karakaya2020simra}).
The idea behind our approach is to take advantage of the size of the data set and use the law of large numbers to obtain robust results without having to calibrate the data and without affecting the results too much by noise.

For this, we consider in the first step each ride individually:
We take the preprocessed ride (which is a single time series of aggregated motion sensor readings without stops plus the GPS trace) and calculate the percentiles (0.2, 0.4, 0.6, 0.8, and 1) of the time series.
For normalization, we then replace all motion sensor readings with values 1 to 5 depending on which interval they fall into, i.e., a motion sensor reading from the interval $(0.2;0.4]$ would be replaced with the value 2.
The intuition behind this is that longer rides are likely to encounter very different surface quality, hence we calculate the relative bumpiness of an area in comparison to the rest of the ride's bumpiness.

In the second step, we use the data from all such normalized rides and, using the GPS trace, map them to a grid of 10m² cells.
For each cell, we hence have a distribution of values 1 to 5.
As a metric for the bumpiness of that cell, we use the average of all values -- e.g., when color-coding a map -- but make other statistical metrics available as well (see, e.g., the distribution function charts in our evaluation section).

\section{Using Road Surface Quality Information}
\label{sec:use}
In this section, we describe how such surface quality results can be used.
In \cref{sec:routing}, we describe how we implement a navigation feature into the SimRa app that uses the road surface quality as an additional parameter in routing.
We then show in \cref{sec:visualization} how we can expose the output data of our pipeline to city administrators.

\subsection{Routing with Surface Quality}%
\begin{figure}
    \centering
    \includegraphics[width=\columnwidth]{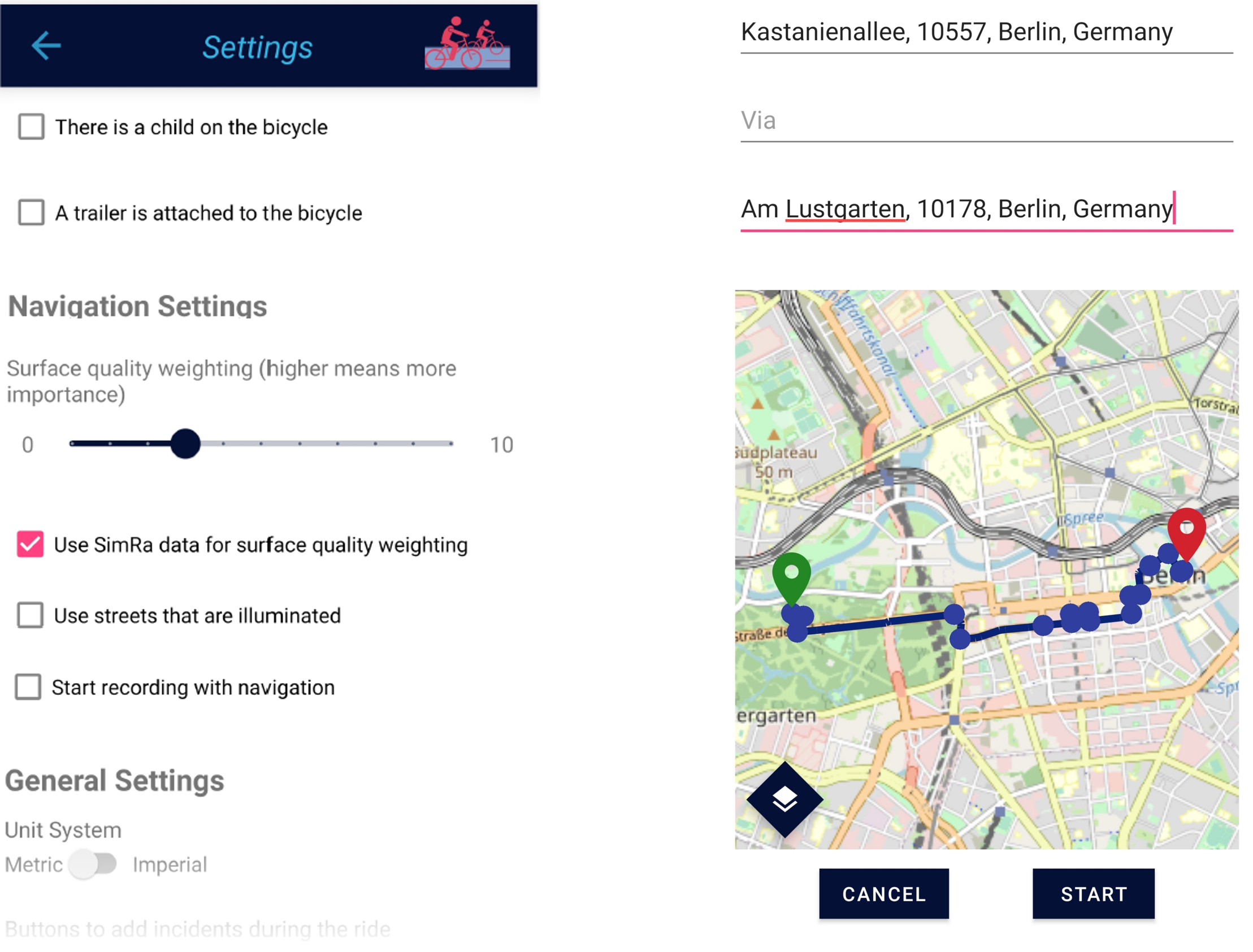}
    \caption{%
With a slider in the settings menu, the weight of the surface quality in the routing can be set.}%
    \label{fig:routing}
\end{figure}

\label{sec:routing}
With the surface quality scores calculated, it is possible to provide a route planning feature, where not only distance and time are considered, but also the surface quality.
The main question here is how much the surface quality should be weighed when calculating the best route from A to B or rather what detour lengths are acceptable.
We decided to give the user the opportunity to influence this factor with the usage of a slider, that can be set between 0 for not considering surface quality in the routing and 10 for the highest importance of the surface quality (\cref{fig:routing}).
We host a modified GraphHopper\footnote{https://github.com/graphhopper/} for routing.
GraphHopper uses edges for streets, that are connected via nodes.
Each has a weight for routing purposes and it is possible to change the weight according to custom data.
This is where we use the surface quality by increasing the weight depending on surface quality and user-specified influence factor.

\subsection{Output Data and Visualization}%
\label{sec:visualization}
Our bicycle road surface quality pipeline creates a GeoJSON file as an output.
It contains the cells with a surface area of 10m² in a grid as \codeword{Features} of the \codeword{geometry} type \codeword{Polygon} with the surface quality score information such as mean, median, standard deviation, number of rides going through the cell and coloring information in the \codeword{properties} key.
 \begin{figure}
    \centering
    \includegraphics[width=\columnwidth]{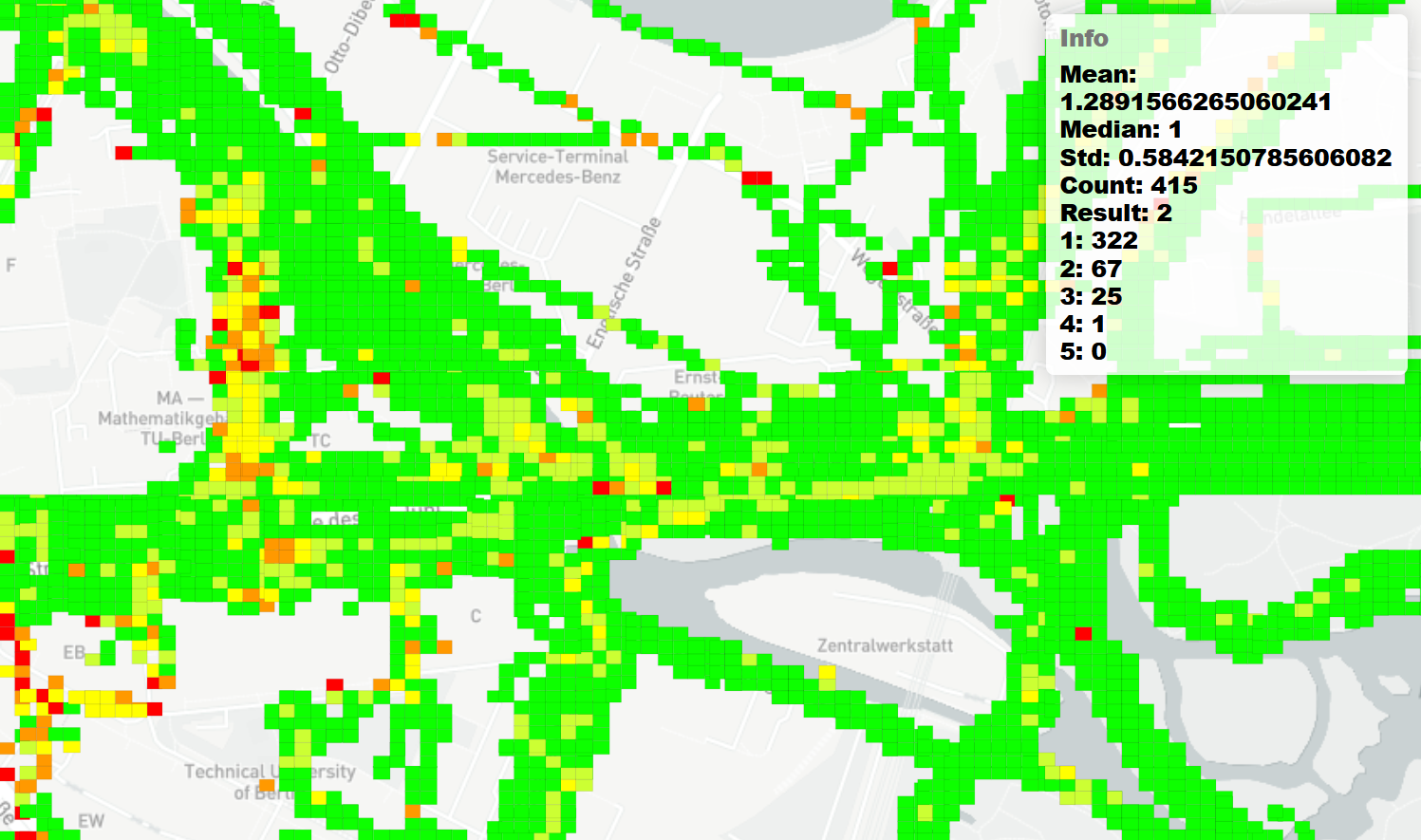}
    \caption{%
A visualization of the output file showing the surface quality of the boxes when hovering over them with the mouse cursor. Color coding is based on the average value.}%
    \label{fig:visualization}
\end{figure}
With such an output file, it is very easy to create a simple visualization\footnote{https://simra-project.github.io/surfaceQuality/Berlin.html}, e.g., with Leaflet\footnote{https://leafletjs.com/}, as depicted in \cref{fig:visualization}.
Using a visualization like this, also non-tech-savvy users can easily monitor the bicycle road surface quality of a large area and take action where needed.

\section{Evaluation}
\label{sec:eval}
In this section, we evaluate the surface quality analysis by comparing the calculated surface quality of selected street segments across Berlin with their surface type in the real world, which we get from OpenStreetMap (OSM).
As data input, we use the existing SimRa datasets~\cite{dataset_simra_set1,dataset_simra_set2,dataset_simra_set3} (almost 90,000 rides with more than 650,000km in total).
Based on the intuition that different surface types will also be partially correlated with surface quality, we randomly picked four spots with different surface types.
To also show the limitations of our approach, we then explored the dataset and manually picked four additional spots where our approach appears to have returned the wrong results.
Each evaluated section has a surface area of 10 m² and to compare them to each other we analyze their mean, median, and standard deviation values.

We first describe the clear results from the first group (\cref{subsec:clear}).
Afterwards, we categorize the additional four ``problem spots'' as mixed results (\cref{subsec:mixed}) or seemingly incorrect results (\cref{subsec:mismatch}).

\subsection{Sections with Clear Results}
\label{subsec:clear}
We evaluate at least one example for each of the following surface types, which are sorted in descending order with regard to their expected surface quality~\cite{titov2019monitoring}: asphalt, flat paving stones, fine gravel, cobblestones.
We chose the sections in a way that (i) asserted that we have sufficient data for them and (ii) to cover all different surface types.
After filtering based on these criteria, we randomly picked four sections.

\begin{figure}
    \centering
    \includegraphics[width=\columnwidth]{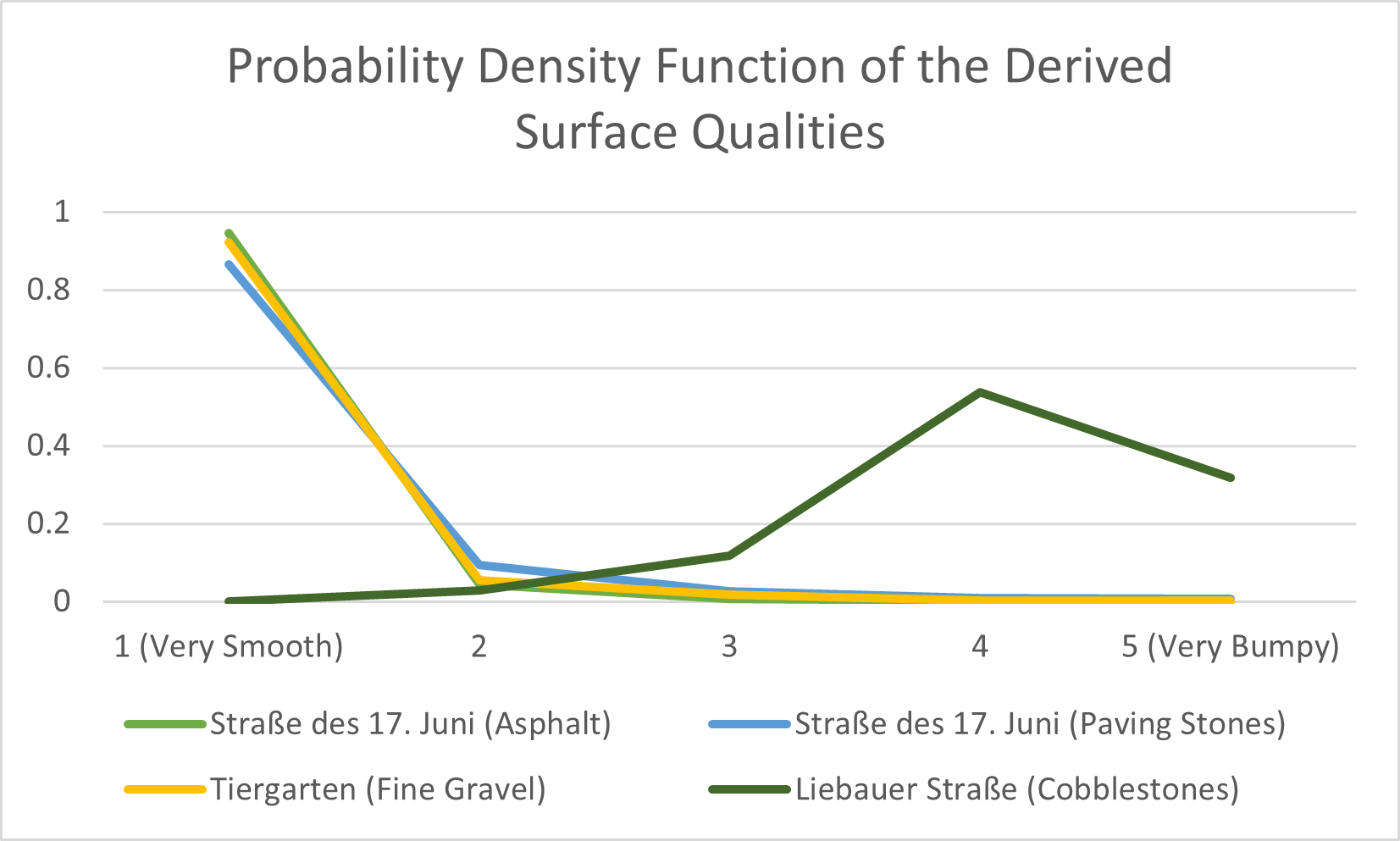}
    \caption{%
        The Probability Density Function of the Derived Surface Qualities shows that these segments have very undisputed road surface quality values, since they are either very good or very bad.
    }%
    \label{fig:clear}
\end{figure}

\cref{tab:clear} and \cref{fig:clear} show the results of the surface quality analysis of the selected sections with very clear and intuitive results.

It can be observed, that the aforementioned list of surface types, which was sorted in descending surface quality order, was ordered correctly.
A newly maintained asphalt section in \textit{Straße des 17. Juni} has a nearly perfect score, which means that its surface quality was in the top 20\% in almost all rides crossing this section.
Followed by that are two sections with flat paving stones and fine gravel as their surface type, which have very similar results.
This means, that both flat paving stones and fine gravel have comparable surface quality in terms of bumpiness from the perspective of a cyclist.
However, it should be noted, that fine gravel can be less favorable in areas with a lot of precipitation (more on that in \cref{sec:disc}).
Not very surprisingly, the \textit{Liebauer Straße} has very bad surface quality scores, since it is paved with cobblestones and has very busy sidewalks with restaurants and cafes, which prevent cyclists from (illegaly) cycling there instead of on the street.

\begin{table}%
\centering
\caption{Surface Quality Analysis Evaluation Results Showing Mean, Median and Standard Deviation of Sections With Clear Results}%
\label{tab:clear}
\resizebox{\columnwidth}{!}{
\begin{tabular}{cccccc}%
\toprule%
Street Name          & Surface       & GPS Location        & Mean & Median & Std. Dev.\\%
\midrule%
\midrule%
Straße des 17. Juni  & Asphalt       & 52.515369,13.630855 & 1.03 & 1      & 0.17\\%
Straße des 17. Juni  & Paving Stones & 52.513501,13.335127 & 1.25 & 1      & 0.5\\%
Tiergarten           & Fine Gravel   & 52.514745,13.34622 & 1.32 & 1      & 1.06\\%
Liebauer Straße      & Cobblestones  & 52.509132,13.453806 & 4.15 & 4      & 1.07\\%
\bottomrule&%
\end{tabular}%
}
\end{table}

\subsection{Sections with Mixed Results}
\label{subsec:mixed}
\begin{figure}
    \centering
    \includegraphics[width=\columnwidth]{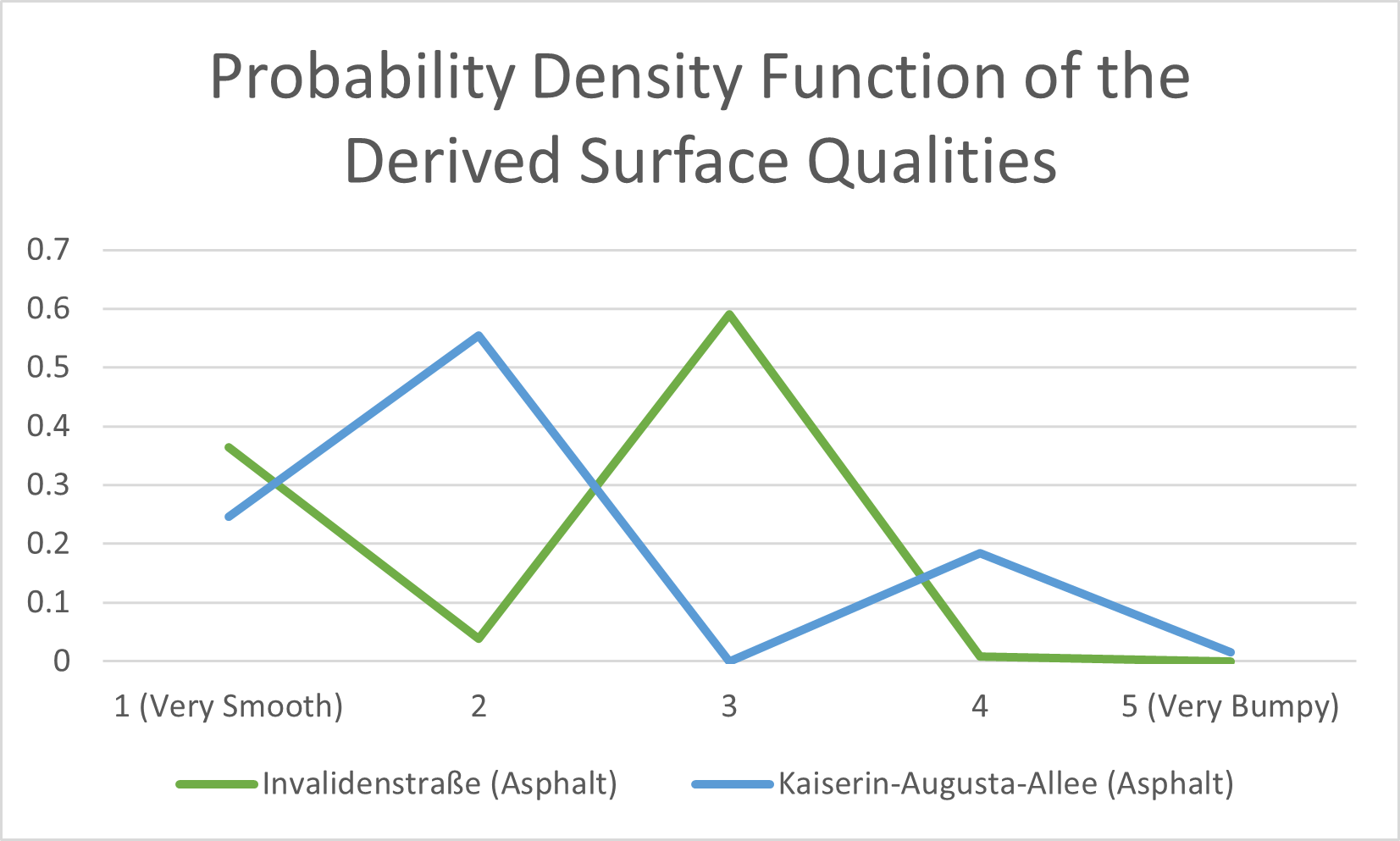}
    \caption{%
        The Probability Density Function of the Derived Surface Qualities shows that these segments have confusing road surface quality values: Depending on the ride, they appear to have either very good or very bad surface quality (multiple peaks).
    }%
    \label{fig:mixed}
\end{figure}

While most results are intuitive, e.g., new bicycle roads paved with asphalt and separated from the motorized vehicle traffic having very good surface conditions, some other sections (see \cref{fig:mixed}) seem confusing at the first glance.

According to the Probability Density Functions (PDFs) of street sections in the \textit{Invalidenstraße} and \textit{Kaiserin-Augusta-Allee}, there seem to be two distinct surface qualities in each section.
A closer look into the specific sections reveal the causes:
 
\begin{table}%
\centering
\caption{Surface Quality Analysis Evaluation Results Showing Mean, Median and Standard Deviation of Sections With Mixed Results}%
\label{tab:mixed}
\resizebox{\columnwidth}{!}{
\begin{tabular}{cccccc}%
\toprule%
Street Name          & Surface       & GPS Location        & Mean & Median & Std. Dev.\\%
\midrule%
\midrule%
Invalidenstraße           & Asphalt   & 52.526236,13.369196 & 2.24 & 3      & 0.96\\%
Kaiserin-Augusta-Allee      & Asphalt  & 52.524429,13.327253 & 2.17 & 2      & 1.05\\%
\bottomrule&%
\end{tabular}%
}
\end{table}

\begin{figure}
    \centering
    \includegraphics[width=\columnwidth]{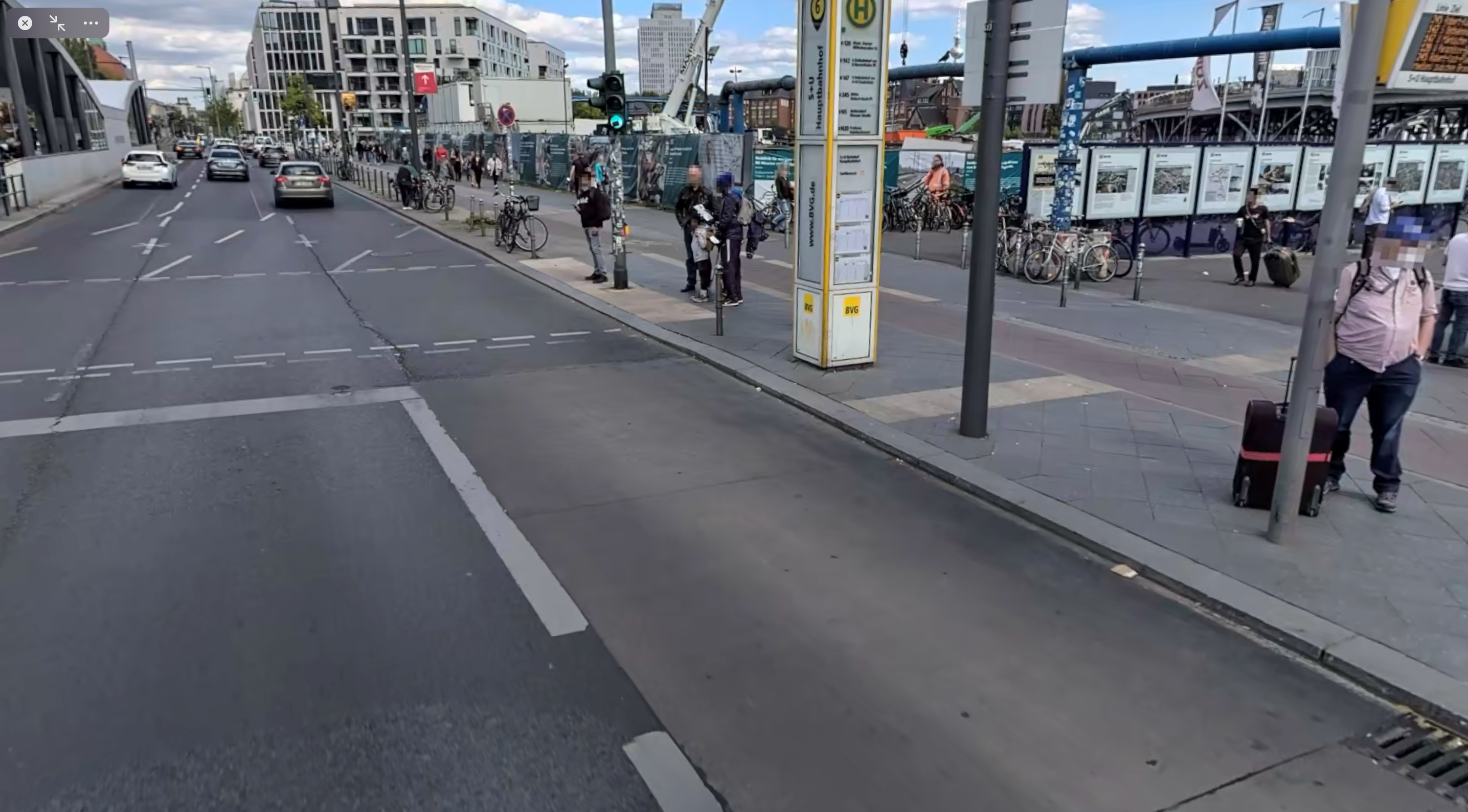}
    \caption{%
        A bus lane, a bus station and a bicycle lane on the sidewalk can create different results in a small area. (Source: Apple Maps)
    }%
    \label{fig:invaliden}
\end{figure}

In \textit{Invalidenstraße}, there are a bus lane, a bus stop, and an on-curb bike lane (see \cref{fig:invaliden}).
Most bus lanes in Berlin can be legally used by cyclists, however, some cyclist may still prefer the relatively bumpy bike lane.
Additionally, when a bus stops at the bus station, cyclists that used the bus lane might overtake the bus on the very smooth car lane to the left. 
These two factors most certainly are the reason for the distinct two-peak distribution of recordings in that spot.

A closer look at the near-miss incident data in \textit{Kaiserin-Augusta-Allee} indicates that people use the bike lane in one direction and the street in the other direction~\cite{dataset_simra_set1,dataset_simra_set2,dataset_simra_set3} as one side seems to be in an unusable condition while the other is acceptable.
Since the bicycle lane is paved with paving stones and the street with asphalt, this leads to different surface quality of the road, depending on which direction the ride was.
Both directions, however, regularly end up in the same 10x10m box as the street is not overly wide, especially considering GPS accuracy.

\subsection{Sections with (Seemingly) Confusing Results}
\label{subsec:mismatch}
\begin{figure}
    \centering
    \includegraphics[width=\columnwidth]{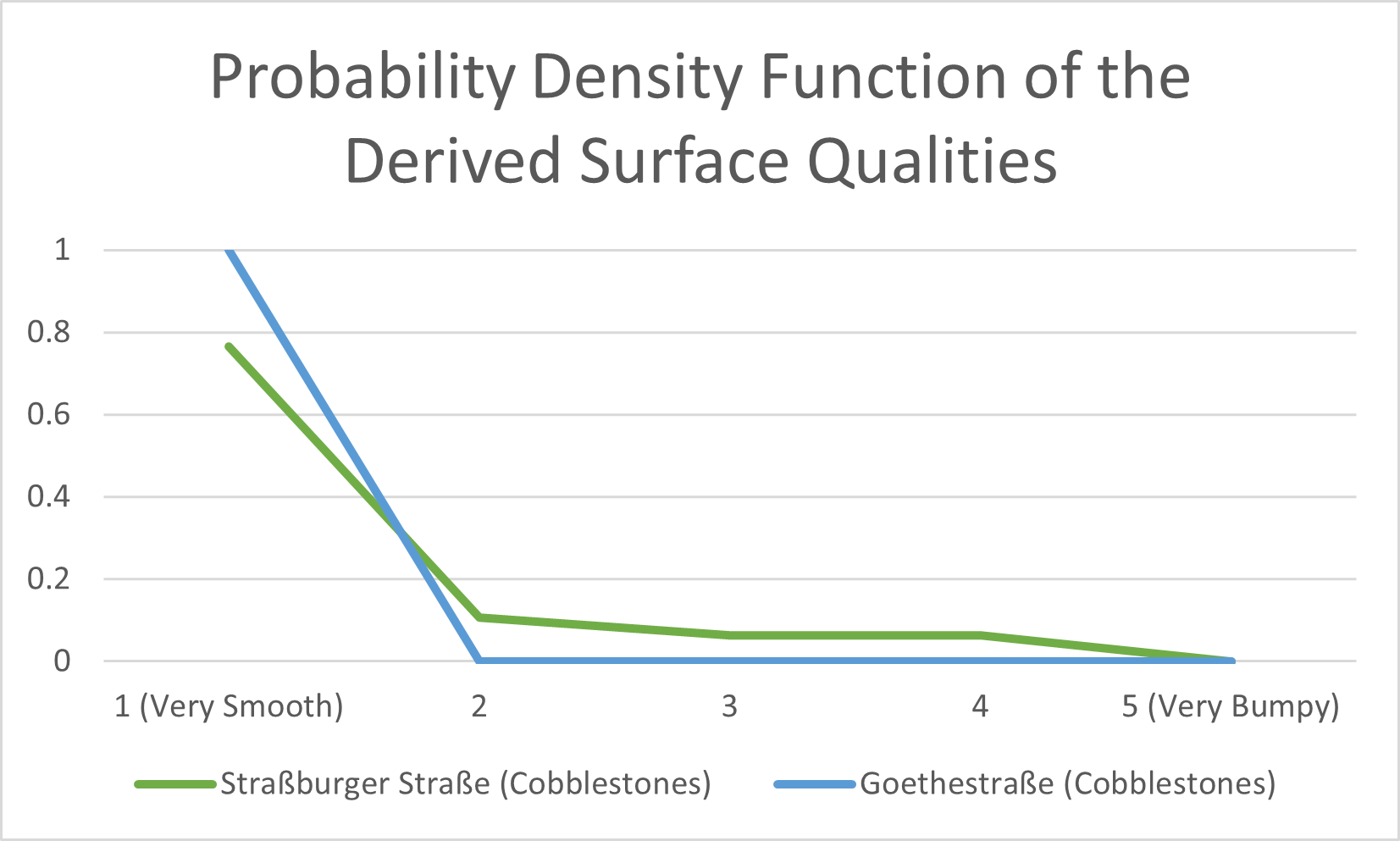}
    \caption{%
        The Probability Density Function of the Derived Surface Qualities shows that these segments have very smooth road surfaces, although they are paved with cobblestones.
    }%
    \label{fig:mismatched}
\end{figure}

There are also sections where the surface quality result seem to be completely wrong, when compared to the OSM surface type labels.
\Cref{tab:mismatch} shows two sections, that are -- according to OSM -- paved with cobblestones and without separate bike lanes, but with very good results, which is confirmed by the PDFs depicted in \cref{fig:mismatched}.

\begin{figure*}[t]
    \centering
    \subfloat[Straßburger Straße]{\includegraphics[width=\columnwidth]{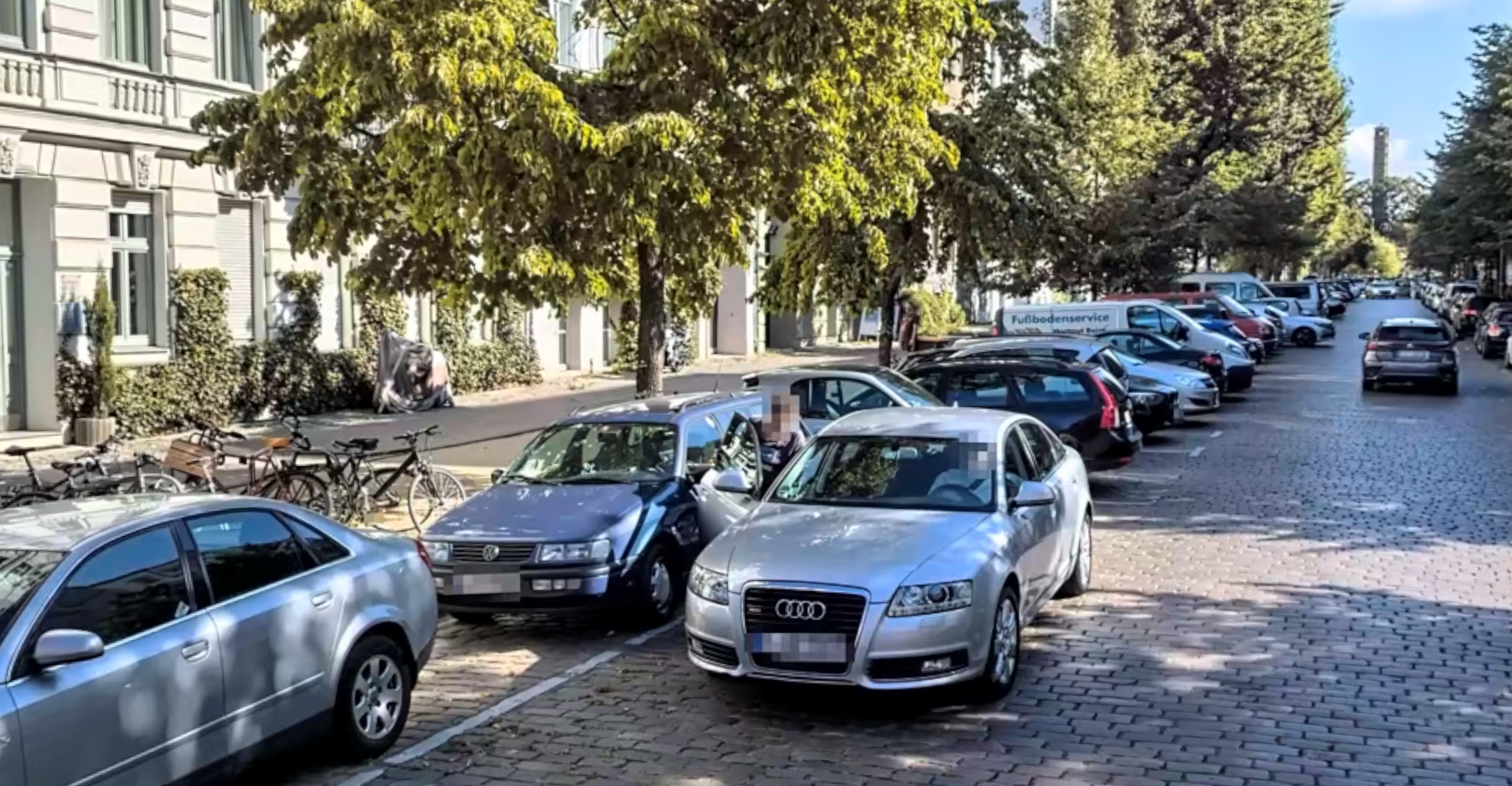}}
    \hfill
    \subfloat[Goethestraße]{\includegraphics[width=\columnwidth]{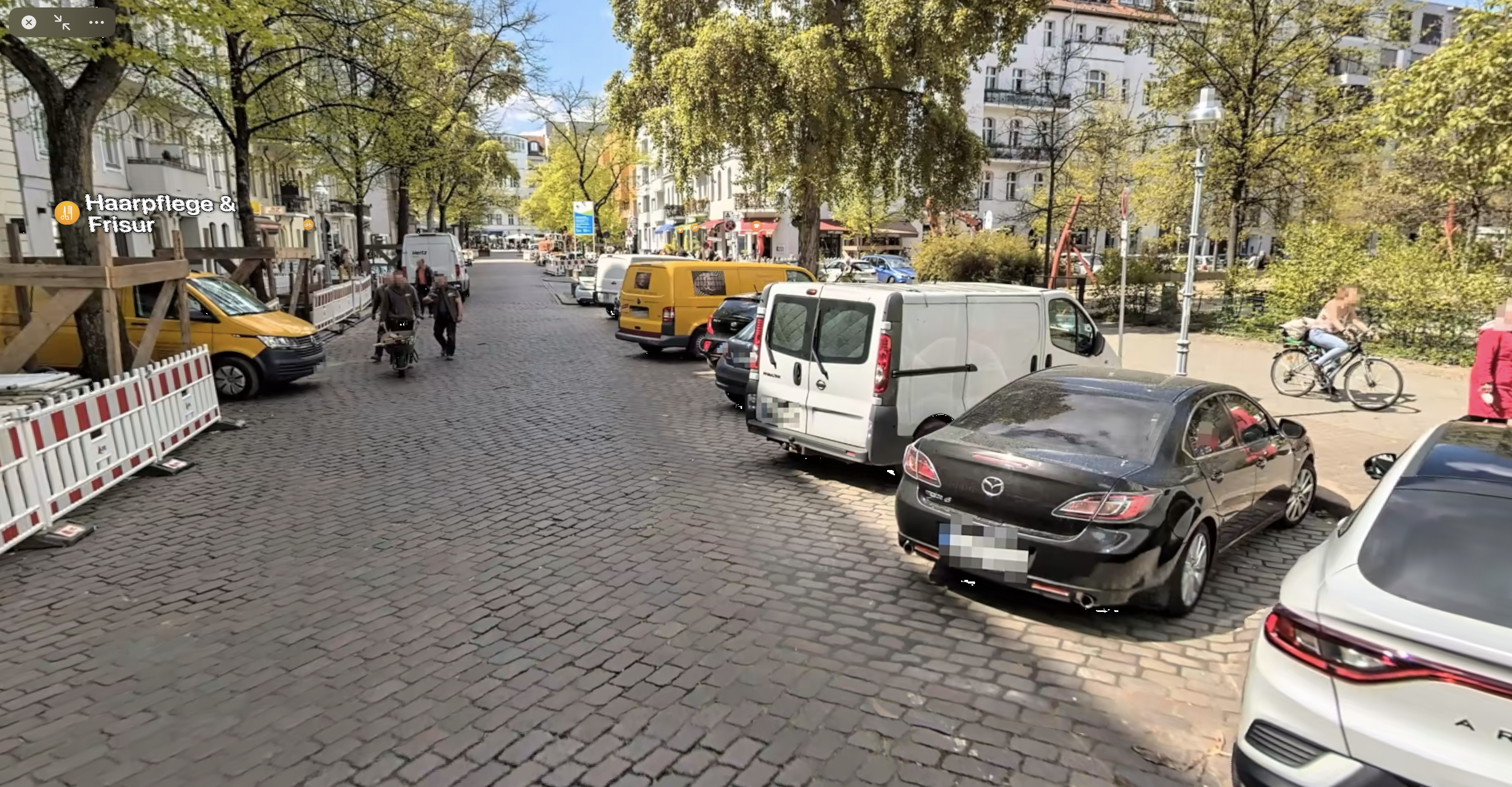}}
    \caption{%
        Straßburger Straße and Goethestraße in Berlin are paved with large cobblestones and the roads are contested by cars that are parking or searching for a parking spot. In contrast, the sidewalks are paved with flat-surfaced paving stones and are quiet due to the absence of shops, restaurants or cafes. (Source: Apple Maps)
    }%
    \label{fig:sidewalk}
    \vspace{-.5em}
\end{figure*}

The possibility, that this section has a wrong label and is in fact, not paved with cobblestones may come to mind.
However, as \cref{fig:sidewalk} reveals, both streets are indeed paved with cobblestones.

\Cref{fig:sidewalk} also shows very likely reasons why these sections have very smooth bicycle rides.
The infrastructural conditions incentivize cyclists to prefer the sidewalks over the actual streets.
First, the cobblestones on the streets make it very unpleasant for cyclists to cycle on them.
Second, the sidewalks are wide, quiet, and paved with large flat-surfaced paving stones.
Third, the streets are contested by parking cars, or by cars searching for a parking spot, service vehicles, construction vehicles, etc. which will often block cyclists and also create safety hazards.
Considering these circumstances, it is not surprising to see good surface quality values in these two and other similar streets as cyclists are likely to (illegaly) use the sidewalks instead.
In fact, photos on Apple Maps (not included in this paper) actually show cyclists using the sidewalk instead.

\begin{table}%
\centering
\caption{Surface Quality Analysis Evaluation Results Showing Mean, Median and Standard Deviation of Sections With (Seemingly) Confusing Results}%
\label{tab:mismatch}
\resizebox{\columnwidth}{!}{
\begin{tabular}{cccccc}%
\toprule%
Street Name        & Surface     & GPS Location        & Mean & Median & Std. Dev.\\%
\midrule%
\midrule%
Straßburger Straße & Cobblestone & 52.532273,13.416521 & 1.43 & 1      & 0.87\\%
Goethestraße       & Cobblestone & 52.508889,13.308333 & 1    & 1      & 0\\%
\bottomrule&%
\end{tabular}%
}
\end{table}

\section{Discussion}
\label{sec:disc}
Overall, the results presented in this paper show, that our approach can derive surface quality information using data recorded in the SimRa app.
Nevertheless, it still has a number of limitations which we discuss in this section.

\subsection{Methodological Challenges}
\label{sec:method}
Since the SimRa dataset consists of crowdsourced data generated by smartphones, the recorded data is very heterogeneous.
This comes from the fact that different smartphone models have different GPS modules and motions sensors, which leads to the problem that a road can be very smooth according to one smartphone and very rough according to another.
This problem is further aggravated by the wide range of different bicycle types, e.g., racing bicycles or mountain bikes, which in turn also heavily influence the vibrations the smartphone can sense.
To solve these problems, we rely on the \textit{law of large numbers} and compare how a section's surface quality was relative to the ride. 

\subsection{Preprocessing in SimRa}
\label{sec:problem_simra}
As input for our surface quality analysis pipeline, we used the SimRa dataset, since it is to our knowledge the only public dataset containing a very high number of anonymized individual rides in a non-aggregated form.
However, it is important to note, that SimRa was developed for identifying near miss incident hotspots in bicycle traffic.
For that, a relatively low accelerometer sampling rate of 50 Hz is used before further reducing the level of detail by calculating the moving average with a window size of 30 and then taking every fifth value.
While with this, it is possible to automatically detect near-miss incidents~\cite{karakaya2022cyclesense}, a higher resolution of data points would allow us to develop a more sophisticated measurement approach.
This would, however, further increase the disk space, memory, battery and bandwidth usage of the SimRa app and as we have shown in \cref{sec:eval}, the surface quality we derive are sufficiently precise for our intended use cases.

\subsection{Behavior of Cyclists}
Due to the GPS inaccuracy, it is impossible to identify whether a cyclist used the actual road or the sidewalk.
For instance, as discussed in the previous section, \cref{fig:sidewalk} showing \textit{Straßburger Straße} in Berlin, would be expected to have poor results but in fact has surprisingly good values.
We believe that this is due to cyclists illegaly using the smooth sidewalk instead of the bumpy road.

Furthermore, cyclists will in practice also avoid the worst potholes and similar bumps if possible, i.e., these will usually be missing from our analysis.

\subsection{Sensor inaccuracy}
The inaccuracy of the GPS sensors~\cite{merry2019smartphone} combined with noise produced by the motion sensors of the smartphones form another limitation of the dataset.
We tried to partially address these limitations with our preprocessing steps but they can, of course, not be fully mitigated.
The only alternative would be using dedicated hardware which, however, will result in significantly less recorded rides due to the adoption barrier.

\subsection{Temporal Influences}
The rides in the SimRa dataset date back up to 2019 and it is possible that the surface quality changed throughout the time.
One reason for that could be that the surface type is changed for example from cobblestones to asphalt or potholes and cracks are repaired.
This would presumably lead to better results after the change, but show up as two-peak distribution in our dataset.
When using this approach in practice, we hence propose to only consider the most recent rides.

Another temporal influence might be the weather.
On rainy, windy or snowy days, the surface quality of the road, especially with wet gravel, can suffer significantly, which is not considered by our surface quality analysis approach.
However, adding this factor to the analysis, would in our dataset reduce the validity of the results, since it would reduce the number of considered rides.

\section{Conclusion}
\label{sec:conclusion}
Cities all over the world aim to increase the modal share of bicycle traffic, e.g., to address emission problems, frequent traffic jams, but also to improve the citizens' health through more daily activity.
Aside from safety, a key influence factor for this is comfort, particularly in the form of the surface quality of cycling infrastructure.
Monitoring the surface quality manually, however, is infeasible due to the dimensions of such infrastructure.

In this paper, we proposed a crowdsensing approach in which cyclists record their daily rides using a smartphone app and the phone's built-in motion sensors.
We proposed a data processing pipeline that starts on the edge (i.e., the phone) and ends in a cloud backend.
Furthermore, we showed that our crowdsensing approach can indeed derive surface quality and implemented two use cases for using such data.

\bibliographystyle{IEEEtran}
\bibliography{bibliography.bib}

\end{document}